\documentclass[24pt]{article}
\usepackage{amssymb}
\usepackage{amsfonts}
\usepackage{amssymb,amsmath}
\usepackage{mathrsfs}
\usepackage{latexsym}
\usepackage{amsmath}
\usepackage{latexsym}
\usepackage[cp1251]{inputenc}
\usepackage{graphicx}
\usepackage{color}

\title{Ground-state properties of a dilute two-dimensional Bose gas}

\author{Volodymyr~Pastukhov\footnote{e-mail: volodyapastukhov@gmail.com}\\ {\small \textit{Ivan Franko
National University of Lviv, Department for Theoretical Physics}}\\
{\small \textit{12 Drahomanov St., Lviv, Ukraine }}}\date{}
\begin{document}
\maketitle
\begin{abstract}
We revisit the problem of the calculation of low-temperature properties for the dilute two-dimensional Bose gas. By using Popov's hydrodynamic approach and perturbation theory on the one-loop level we recover not only the known expansion for the ground-state energy but also calculate for the first time the condensate density and Tan's contact.
\end{abstract}
\section{Introduction}
It is well-known that the role of both quantum and thermal fluctuations has a crucial impact on the properties of many-body systems when the spatial dimension decreases. For interacting fermions their presence leads to a very intriguing paradigm shift in the theoretical descriptions from the Landau Fermi Liquid in higher dimensions to the Luttinger Liquid approach, when the dimensionality is equal to unity. In the case of bosons the situation is somewhat simpler because there is a possibility to capture the whole variety of phases of a dilute Bose gas in any dimension within a single approach at least at very low temperatures. Of course, the richest phase diagram is observed in the two-dimensional (2D) geometry (see, for review \cite{Posazhennikova,Hadzibabic}). In fact, this system undergoes two phase transitions, namely the Bose-Einstein condensation at absolute zero and vanishing of the off-diagonal long-range order at finite temperatures. Recent interest to 2D bosonic systems is stimulated by experimental observation \cite{Hadzibabic_exp,Clade,Yefsah,Desbuquois} of superfluidity accompanied by the emergence of long-range coherence in the quasi-two-dimensional ultracold gases. Particularly the impact of three-body effects on the properties of system was studied theoretically in Ref.~\cite{Mashayekhi}, while the magnitude of finite-range corrections to the low-temperature thermodynamics is still debated \cite{Salasnich,Beane_17}. Nevertheless, the first attempts to describe two-dimensional Bose gases at zero \cite{Schick} and non-zero \cite{Popov_72} temperatures go back to early 1970s, when it was shown that the thermodynamics in a dilute limit is controlled by an intrinsic small parameter $1/|\ln na^2|$ (with $a$ and $n$ being $s$-wave scattering length and density of the system, respectively). A modern view on the structure of this expansion was formed in Ref.~\cite{Lozovik}, reproduced extensively by various analytical approaches \cite{Kolomeisky,Ovchinnikov,Cherny,Andersen,Mora_03,Mora09,Astrakharchik_10,Beane_10,Pricoupenko,Abdelaali}, generally confirmed by the quantum Monte Carlo simulations \cite{Astrakharchik_09}, and tested experimentally \cite{Ha}. 
The impact of repulsive interparticle interaction on the behavior of condensate in 2D Bose systems was studied variationally \cite{Mazzanti}, numerically \cite{Pilati}, and by means of the leading-order-auxiliary-field theory \cite{Chien}. However, the analytical evaluation of the first beyond Bogoliubov correction to a condensate depletion of dilute gases within conventional theoretical approaches is very cumbersome. In the present article we apply the Popov hydrodynamic description to resolve this difficulty.

\section{Method description}
We consider a model of $N$ bosons loaded in a square of large area $\mathcal{A}$ with the periodic boundary conditions imposed. All particles have the same mass $m$ and the interaction between them is assumed to be pairwise with the two-body potential $\Phi({\bf r})$ characterized by some coupling constant which in turn is dependent on the effective range parameter. The specific form of this interaction is not important for our calculations because all final formulas will be written in terms of the universal $s$-wave scattering length (see \cite{Volosniev,Konietin} for details). The method presented below, nevertheless, is also applicable to the systems with three-body, four-body and higher-order interparticle interactions as well. In the following we adopt the imaginary-time path-integral formulation \cite{Popov}. The Euclidean action of the model then reads
\begin{align}\label{S}
&S=\int dx \,\psi^*(x)\left\{\frac{\partial}{\partial\tau}+\frac{\hbar^2\nabla^2}{2m}+\mu
\right\}\psi(x)\nonumber\\
&-\frac{1}{2}\int dx\int dx'\Phi(x-x')|\psi(x)|^2|\psi(x')|^2,
\end{align}
where $\Phi(x)=\delta(\tau)\Phi({\bf r})$, $\mu$ is the chemical potential and integration over $x=(\tau, {\bf r})$ is carried out in the $(2+1)$-``volume'' $\beta \mathcal{A}$ (here $\beta$ is the inverse temperature). Introducing auxiliary parameter $\Lambda$ such that $\hbar^2\Lambda^2/m\gg \mu$ and applying Popov's procedure of separation of the so-called ``slowly'' and ``rapidly'' varying fields with the subsequent functional integration over
the rapid one we are in position to obtain the effective action governing the properties of the system. For dilute Bose systems at zero temperature in the limit of vanishing interaction range the structure of action is similar to Eq.~(\ref{S}) 
\begin{eqnarray}\label{S_eff}
S_{{\rm eff}}=\int dx \,\psi^*(x)\left\{\frac{\partial}{\partial\tau}+\frac{\hbar^2\nabla^2}{2m}+\mu
\right\}\psi(x)-\frac{1}{2}\int dx\mathcal{T}|\psi(x)|^4,
\end{eqnarray} 
but with the effective local two-body potential $\Phi_{\rm eff}({\bf r})=\mathcal{T}\delta({\bf r})$, where the $\Lambda$-dependent coupling constant
\begin{eqnarray}\label{Tau}
\mathcal{T}^{-1}=t^{-1}-\frac{1}{ \mathcal{A}}\sum_{|{\bf k}|<\Lambda}\frac{1}{2\varepsilon_k+\epsilon},
\end{eqnarray}
is related to the $s$-wave scattering length $a$ through $t^{-1}=\frac{m}{4\pi\hbar^2}\ln\left |\frac{\epsilon_b}{\epsilon}\right|$. Here the two-particle binding energy reads $\epsilon_b=-\frac{4e^{-2\gamma}\hbar^2}{ma^2}$
($\gamma=0.57721\ldots$ is the Euler-Mascheroni constant) and for latter convenience notation for the free-particle dispersion $\varepsilon_k=\hbar^2k^2/2m$ and auxiliary energy scale $\epsilon$ are also introduced. In the dilute limit a dummy parameter $\epsilon>0$ is of the order of the system chemical potential $\mu$ but the final results are unaffected by its precise value.

The Mermin–Wagner theorem states that phases with spontaneously broken continuous symmetries are thermodynamically forbidden at finite temperatures when dimensionality of space is less or equal two. In context of many-boson theory it means that there is no condensate in such systems and of course the use of conventional Bogoliubov approach is questionable in this case. Furthermore even in higher dimensions ($>2$) the perturbation theory for Bose systems with separated condensate is complicated by the infrared divergences. Therefore, in order to avoid all these difficulties we adopt the phase-density representation for the bosonic fields $\psi^*(x)=\sqrt{n(x)}e^{-i\varphi(x)}$, $\psi(x)=\sqrt{n(x)}e^{i\varphi(x)}$. This change of variables in the path-integral leads us to the hydrodynamic action \cite{Pastukhov_InfraredStr}, which correctly determines the low-temperature properties of dilute Bose systems in any dimension
\begin{align}\label{S_hydro}
S_{{\rm eff}}=\int dx\, \left\{n(x)i\partial_{\tau}\varphi(x)+\mu n(x)-\frac{1}{2}\mathcal{T}n^2(x)\right.\nonumber\\
\left.-\frac{\hbar^2}{2m}n(x)[\nabla\varphi(x)]^2-\frac{\hbar^2}{8m}\frac{[\nabla n(x)]^2}{n(x)}\right\},
\end{align}
and the Bose condensate (if it exists, of course) is manifested as a off-diagonal long-range order of the one-body density matrix $\langle \psi^*(x) \psi(x')\rangle|_{\tau'\to \tau, |{\bf r}-{\bf r}'|\to \infty}=n_0$. Then making use of the Fourier transform
\begin{eqnarray}\label{n_varphi}
n(x)=n+\frac{1}{\sqrt{\beta\mathcal{A}}}\sum_{K}e^{iKx}n_{K}, \ \
\varphi(x)=\frac{1}{\sqrt{\beta\mathcal{A}}}\sum_{K}e^{iKx}\varphi_{K},
\end{eqnarray}
where  $K=(\omega_k, {\bf
k})$ stands for the bosonic Matsubara frequency $\omega_k$ and two-dimensional wave-vector ${\bf k}$ (recall that $k\le\Lambda$ and ${\bf k}\neq 0$), we can expand the action (\ref{S_hydro}) to obtain
\begin{eqnarray}\label{S_F}
S_{{\rm eff}}=\beta\mathcal{A}n\mu-\frac{1}{2}\beta\mathcal{A}\mathcal{T}n^2-\frac{1}{2}\sum_{K}
\left\{\vphantom{\left[\frac{\varepsilon_k}{2 n}
	+\mathcal{T}\right]}\omega_k\varphi_K n_{-K}-\omega_k\varphi_{-K}n_{K}
\right.\nonumber\\
\left.+2n\varepsilon_k
\varphi_{K}\varphi_{-K}+\left[\frac{\varepsilon_k}{2 n}
+\mathcal{T}\right]n_{K}n_{-K}\right\}\nonumber\\
+\frac{1}{2\sqrt{\beta\mathcal{A}}}\sum_{K,
	Q}\frac{\hbar^2}{m}{\bf
	kq}n_{-K-Q}\varphi_{K}\varphi_{Q}\nonumber\\
+\frac{1}{3!\sqrt{\beta
		\mathcal{A}}}\sum_{K+Q+P=0}\frac{1}{4n^2}(\varepsilon_k+\varepsilon_q+\varepsilon_p)n_{K}n_{Q}n_{P}\nonumber\\
-\frac{1}{8\beta
	\mathcal{A}}\sum_{K,Q}\frac{1}{n^3}(\varepsilon_k+\varepsilon_q)n_{K}n_{-K}n_{Q}n_{-Q}+\dots,
\end{eqnarray}
where only terms relevant for our second-order beyond-mean-field calculations are presented. The steepest descent method together with the thermodynamic relation $-\partial \Omega/\partial\mu=N$ for the grand-canonical potential identify $n$ \cite{Pastukhov_InfraredStr} to be the density of the system. Thus, the hydrodynamic approach allows to proceed with further consideration in the canonical ensemble and directly calculate the free-energy of the Bose gas perturbatively.

\subsection{Condensate density}
In the following we consider the zero-temperature limit only and having calculated the ground state energy $E$ explicitly we are in position to obtain another observables by using various variational theorems. For instance, the distribution function of particles with non-zero momentum can be derived as follows
\begin{eqnarray}\label{N_k}
	N_k=\left(\frac{\delta \Omega}{\delta \varepsilon_k}\right)_{\mu, \mathcal{T}}.
\end{eqnarray}
A counterpart of the above formula suitable for use in the hydrodynamic approach is $N_k=\left(\frac{\delta E}{\delta \varepsilon_k}\right)_{n,\mathcal{T}}$. Then for the number of particles in the Bose condensate we naturally find
\begin{eqnarray}\label{N_0}
N_0=N-\sum_{|{\bf k}|<\Lambda}N_k.
\end{eqnarray}
It should be noted that the above formula has a distinct meaning only at absolute zero, where the condensate in a two-dimensional Bose gas really exists. This fact provides that the calculation of $N_0$ is only important from the methodological point of view because in practical realization of 2D systems the thermal fluctuations at any finite temperatures totally destroy the Bose condensate.

\subsection{Tan's energy relation and contact}
The ground-state energy of the system can be also obtained as a sum of average kinetic $K$ and potential $\Phi$ energies. For the calculation of these quantities we will use the Hellmann--Feynman-type theorems, namely $K=-m\frac{\partial\Omega}{\partial m}$ and $\Phi=-t\frac{\partial\Omega}{\partial t}$. Then the simple evaluation of the energy density with action (\ref{S_eff}) yields
\begin{eqnarray}\label{E_T}
\frac{E}{\mathcal{A}}=\frac{{\mathcal{T}\,}^2\langle|\psi(x)|^4\rangle}{2t}+\frac{1}{\mathcal{A}}\sum_{|{\bf k}|	\le \Lambda}\varepsilon_k\left\{N_k-\frac{{\mathcal{T}\,}^2\langle|\psi(x)|^4\rangle}{2\varepsilon_k(2\varepsilon_k+\epsilon)}\right\}.
\end{eqnarray}
This equation is a two-dimensional bosonic counterpart \cite{Combescot,Valiente} of the Tan energy relation \cite{Tan_1,Braaten} originally obtained for 3D systems. The large-$k$ tail of particle distribution $N_k$ determines the so-called Tan's contact $\mathcal{C}=\left(\frac{m\mathcal{T}}{\hbar^2}\right)^2 \langle|\psi(x)|^4\rangle$. This result particularly suggests that the product
${\mathcal{T}\,}^2\langle|\psi(x)|^4\rangle$ remains finite during the renormalization scheme (\ref{Tau}) in every order of expansion over $t$. 

The contact can be explicitly written in a form convenient for calculations in the hydrodynamic approach
\begin{eqnarray}\label{C_h}
\mathcal{C}=\left(\frac{m\mathcal{T}}{\hbar^2}\right)^2 \left\{n^2+\frac{1}{\mathcal{A}}\sum_{|{\bf k}|<\Lambda}(S_k-n)\right\},
\end{eqnarray}
where the pair structure factor of the system reads $S_k=\frac{1}{\beta}\sum_{\omega_k}\langle n_Kn_{-K}\rangle $. The evaluation of $S_k$ can be performed by using another variational theorem. Restoring for a moment the dependence of an effective coupling constant  on the wave-vector $\mathcal{T}\to \mathcal{T}_k$ in formula for action (\ref{S_F}) and mentioning that 
\begin{eqnarray}\label{S_k}
S_k-n=2\left(\frac{\delta \Omega}{\delta \mathcal{T}_k}\right)_{\mu}=2\left(\frac{\delta E}{\delta \mathcal{T}_k}\right)_{n},
\end{eqnarray}
we are in position to calculate not only contact $\mathcal{C}$ but also diagonal elements of the two-body density matrix (the pair distribution function) $\langle |\psi(x)|^2 |\psi(x')|^2\rangle|_{\tau'\to \tau}$.

\section{Perturbation theory}
In general, the path integral with action (\ref{S_F}) cannot be calculated exactly, but for a dilute system one may use perturbation theory in terms of gas parameter $na^2$. Absence of infrared divergences guarantees that this expansion is well-defined and by taking into account first three terms of the Fourier-transformed action $S_{\rm eff}$ we recover the celebrated Bogoliubov approximation. The remaining explicitly written terms in Eq.~(\ref{S_F}) contribute to the free energy of system on the one-loop level (for details of the perturbation theory applied to hydrodynamic action $S_{\textrm{eff}}$ see \cite{Popov,Pastukhov_q2D,Pastukhov_twocomp}). These second-order beyond-mean-field correction has a simple diagrammatic representation (see Fig.~1). 
\begin{figure}[h!]
	\centerline{\includegraphics
		[width=1.0\textwidth,clip,angle=-0]{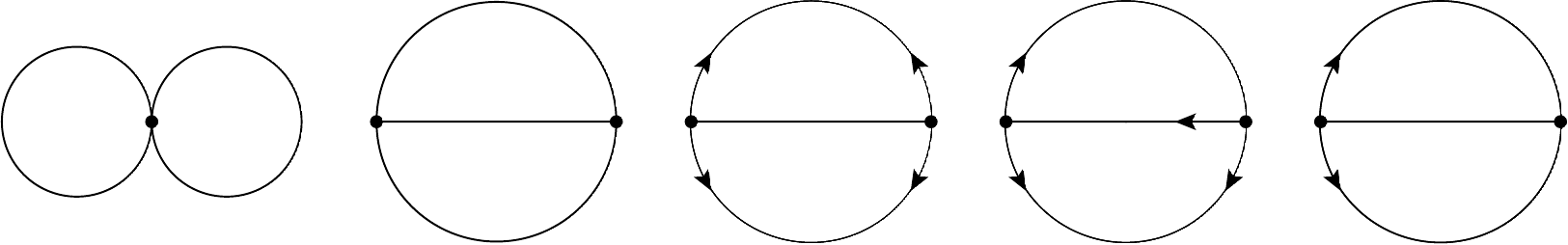}}
	\caption{Diagrams contributing to the beyond Bogoliubov free energy. Lines with and without arrows represent phase and density fluctuation fields, respectively. Dots stand for various bare vertex functions originating from the hydrodynamic action (\ref{S_F}).}
\end{figure}

Tedious calculations in the zero-temperature limit leave us with the following result
\begin{eqnarray}\label{E}
E=E_B+\Delta E,
\end{eqnarray}
where the mean-field term plus the Bogoliubov correction read
\begin{eqnarray}\label{E_B}
E_B=\frac{1}{2}\mathcal{A}n^2\mathcal{T}+\frac{1}{2}\sum_{|{\bf k}|\le \Lambda}\left(E_k-\varepsilon_k-n\mathcal{T}\right),
\end{eqnarray}
and the one-loop contribution is given by
\begin{align}\label{Delta_E}
&\Delta E=-\frac{1}{96N}\sum_{{\bf k}+{\bf q}+{\bf s}=0}\frac{1}{\alpha_q\alpha_k\alpha_s}
\frac{f^2(k,q,s)}{E_q+E_k+E_s}\nonumber\\
&+\frac{1}{24N}\sum_{{\bf k}+{\bf q}+{\bf s}=0}(\varepsilon_q+\varepsilon_k+\varepsilon_s)\left(1-\frac{1}{\alpha_q}\right)\left(1-\frac{1}{\alpha_k}\right)\left(1-\frac{1}{\alpha_s}\right),
\end{align}
with $E_k=\sqrt{\hbar^4k^4/4m^2+n\mathcal{T}\hbar^2k^2/m}$, ($\alpha_k=E_k/\varepsilon_k$) being Bogoliubov's spectrum and
\begin{eqnarray}\label{f}
f(k,q,s)=(\hbar^2/m)\left[{\bf k}{\bf q}(\alpha_k-1)(\alpha_q-1)+{\textrm{perm.}}\right],
\end{eqnarray}
denoting symmetric function of its arguments. The obtained formulas (\ref{E}), (\ref{Delta_E}) constitute the central result of this paper and will be used below for the evaluation of the ground-state energy, condensate depletion and contact of a two-dimensional dilute Bose gas. But in order to obtain some reasonable value for the energy of a system at this stage of calculations we have to replace the $\Lambda$-dependent coupling constant $\mathcal{T}$ via its expansion over $t$. Then the upper integration limit in Eqs.~(\ref{E_B}),~(\ref{Delta_E}) can be tended to infinity. This renormalization procedure should be also performed during the calculations of condensate density: first by using Eqs.~(\ref{E})-(\ref{Delta_E}) we have to evaluate particle distribution $N_k$ at fixed $\mathcal{T}$, then substitute $\mathcal{T}$ in terms of $t$ in the obtained result and finally calculate the integrals determining condensate density (\ref{N_0}). The regularization is needed during the calculations of Tan's contact. Again, after the evaluation of derivative $\left(\frac{\delta E}{\delta \mathcal{T}_k}\right)_{n}$ with the following substitution in formula (\ref{C_h}), we have to replace $\mathcal{T}$ (dropping subscript $k$) via its series expansion over $t$, and then carry out all integrals.

\section{Results}

Let us start first with the calculation of the ground-state energy. It is known that every perturbation treatment \cite{Cherny,Andersen,Mora_03,Mazzanti,Mora09} usually gives a result as an expansion over the inherent small parameter $1/|\ln na^2|$. The same, of course, regards our hydrodynamic approach. On the other hand, from the results of Monte Carlo simulations \cite{Astrakharchik_10,Astrakharchik_09} we find out that the correct form of energy is the following
\begin{eqnarray}\label{E_res}
\frac{E}{\mathcal{A}}=\frac{2\pi \hbar^2n^2/m}{|\ln na^2|+\ln|\ln na^2|+C^E_1+\frac{\ln|\ln na^2|+C^E_2}{|\ln na^2|}+\ldots}.
\end{eqnarray}
The first constant is the same in every approximate treatment $C^E_1=-\ln \pi-2\gamma-1/2$, but value of $C^E_2$ may vary from one approach to another. The numerical fit gives $C^E_2\simeq -0.05$ \cite{Astrakharchik_10,Astrakharchik_09}. Our estimation looks like $C^E_2=-\ln \pi-2\gamma-3/4-\textrm{const}_E$, where the introduced constant is equal to $\textrm{const}_E\simeq-2.54$ and given by integrals in Appendix. In fact our result for $C^E_2\simeq -0.51$ reproduces the known value obtained earlier in Ref.~\cite{Mora09} for a model with condensate. This consistency of two different calculation techniques is very important for justification of the hydrodynamic approach itself and, of course, for the computations performed below.

A very similar series expansion in powers of the presented above inverse logarithms was also obtained for the condensate density of two-dimensional bosons. Likewise to the energy calculations, assuming a fraction structure for the depletion of condensate we obtained
\begin{eqnarray}\label{N_0_res}
\frac{N_0}{N}=1-\frac{1}{|\ln na^2|+\ln|\ln na^2|+C^{N_0}_1+\ldots},
\end{eqnarray}
where $C^{N_0}_1=-\ln \pi-2\gamma-1-\textrm{const}_{N_0}$ and numerically evaluated $\textrm{const}_{N_0}\simeq 1.63$ (for integrals determining the latter constant see Appendix). Note that in contrast to the three-dimensional case, where both the second-order energy \cite{Hugenholtz} and condensate corrections \cite{Schakel} remain to be logarithmically divergent even after regularization procedure (i.e., dependent on the range of a two-body potential), the thermodynamic properties of a 2D Bose gas are universal at this stage of calculations.

The complicated structure of Eq.~(\ref{E_res}) and the existence of an exact identity relating energy and contact \cite{Werner} in two dimensions
\begin{eqnarray}
\frac{1}{\mathcal{A}}\frac{\partial E}{\partial \ln a}=\frac{\hbar^2 \mathcal{C}}{4\pi m},
\end{eqnarray}
forced us to write down the series expansion for $\mathcal{C}$ coming from Eq.~(\ref{C_h}) in the explicit form
\begin{align}\label{C_res}
&\mathcal{C}=\left(\frac{4\pi n}{\ln na^2}\right)^2\left\{1-2\frac{\ln|\ln na^2|+C^{\mathcal{C}}_1}{|\ln na^2|}\right.\nonumber\\
&\left.+\frac{[\ln|\ln na^2|+C^{\mathcal{C}}_1][3\ln|\ln na^2|+3C^{\mathcal{C}}_1-2]+\textrm{const}_{\mathcal{C}}}{\ln^2 na^2}\right\},
\end{align}
where $C^{\mathcal{C}}_1=-\ln \pi-2\gamma-1$ and $\textrm{const}_{\mathcal{C}}$ should be equal to $3\textrm{const}_{E}$ in accordance with the above exact identity. The straightforward numerical calculations of integrals from Appendix confirm this suggestion.


\section{Conclusions}
In conclusion, we have analyzed by means of path-integral approach the thermodynamics of a two-dimensional Bose gas at zero temperature. Starting from the microscopic model that explicitly takes into account only pairwise interaction between bosons and applying original Popov's procedure we have obtained the hydrodynamic action governing the properties of a system. Working on the one-loop level we have calculated the second-order beyond-mean-field ground-state energy correction and demonstrated the consistency of our result with that of the conventional Bogoliubov approach. The latter observation proves that the path-integral formulation in terms of phase and density fluctuations is useful not only in the description of infrared physics of Bose systems, but also is capable for the calculations of thermodynamic properties. Particularly it allowed us to obtain the next to leading order term in the series expansion for the condensate depletion of a 2D dilute Bose gas, which to our knowledge has not been reported yet, and to provide a simple derivation of Tan's energy relation in two dimensions.

\begin{center}
	{\bf Acknowledgements}
\end{center}
We thank Prof.~A.~Rovenchak for stimulating discussions. This work was partly supported by Project FF-30F (No.~0116U001539) from the Ministry of Education and Science of Ukraine.

\section{Appendix}
In this section we present some details of calculation of integrals determining constants in the ground-state energy (\ref{E_res}), condensate depletion (\ref{N_0_res}) and contact (\ref{C_res}). We will not stop on the evaluation of variational derivatives $\left(\frac{\delta E}{\delta \varepsilon_k}\right)_{n,\mathcal{T}}$, $\left(\frac{\delta E}{\delta \mathcal{T}_k}\right)_{n}$ with the following integration in ${\bf k}$-space and only give the dimensionless expressions written in terms of triple integrals. After elimination of the explicit dependence on cut-off parameter $\Lambda$ for $\textrm{const}_E$ we obtained
\begin{align}
\textrm{const}_E=\frac{32}{\pi}\int^{\infty}_{0}dk\left(1-\frac{k}{\sqrt{k^2+1}}\right)
\int^{\infty}_{0}dq\left(1-\frac{q}{\sqrt{q^2+1}}\right)\nonumber\\
\times\int^{k+q}_{|k-q|}ds s\left\{1-\frac{(s^2-k^2-q^2)^2}{4k^2q^2}\right\}^{-1/2}\left(s^2-\frac{s^3}{\sqrt{s^2+1}}-\frac{1}{2}\right)\nonumber\\
-\frac{32}{3\pi}\int^{\infty}_{0}\frac{dkk}{\sqrt{k^2+1}}
\int^{\infty}_{0}\frac{dqq}{\sqrt{q^2+1}}
\int^{k+q}_{|k-q|}ds s\left\{1-\frac{(s^2-k^2-q^2)^2}{4k^2q^2}\right\}^{-1/2}\nonumber\\
\times\frac{s}{\sqrt{s^2+1}}\frac{{\tilde{f}}^2(k,q,s)}{k\sqrt{k^2+1}+q\sqrt{q^2+1}+s\sqrt{s^2+1}},
\end{align}
here and below
\begin{align*}
{\tilde{f}}(k,q,s)=\frac{s^2-k^2-q^2}{2}\left(\sqrt{1+1/k^2}-1\right)\left(\sqrt{1+1/q^2}-1\right)+\textrm{perm.}
\end{align*}
The second-order correction to condensate density is determined by the following constant
\begin{align}
\textrm{const}_{N_0}=\frac{8}{\pi}\int^{\infty}_{0}dk\left(1-\frac{k}{\sqrt{k^2+1}}\right)
\int^{\infty}_{0}dq\left(1-\frac{q}{\sqrt{q^2+1}}\right)\nonumber\\
\times\int^{k+q}_{|k-q|}ds s\left\{1-\frac{(s^2-k^2-q^2)^2}{4k^2q^2}\right\}^{-1/2}\left(1-\frac{s}{\sqrt{s^2+1}}-\frac{s}{2(s^2+1)^{3/2}}\right)\nonumber\\
-\frac{8}{\pi}\int^{\infty}_{0}dk\left(1-\frac{k}{\sqrt{k^2+1}}\right)
\int^{\infty}_{0}dq\left(q^2-\frac{q^3}{\sqrt{q^2+1}}-\frac{1}{2}\right)\nonumber\\
\times\int^{k+q}_{|k-q|}ds\left\{1-\frac{(s^2-k^2-q^2)^2}{4k^2q^2}\right\}^{-1/2}\frac{1}{(s^2+1)^{3/2}}\nonumber\\
-\frac{4}{\pi}\int^{\infty}_{0}\frac{dkk}{\sqrt{k^2+1}}
\int^{\infty}_{0}\frac{dqq}{\sqrt{q^2+1}}
\int^{k+q}_{|k-q|}ds \left\{1-\frac{(s^2-k^2-q^2)^2}{4k^2q^2}\right\}^{-1/2}\nonumber\\
\times\frac{\partial}{\partial s}\frac{s}{\sqrt{s^2+1}}\frac{{\tilde{f}}^2(k,q,s)}{k\sqrt{k^2+1}+q\sqrt{q^2+1}+s\sqrt{s^2+1}}.
\end{align}
Finally, $\textrm{const}_{\mathcal{C}}$ after some rearrangements is given by
\begin{align}
\textrm{const}_{\mathcal{C}}=-\frac{16}{\pi}\int^{\infty}_{0}dk\left(1-\frac{k}{\sqrt{k^2+1}}\right)
\int^{\infty}_{0}dq\left(1-\frac{q}{\sqrt{q^2+1}}\right)\nonumber\\
\times\int^{k+q}_{|k-q|}ds s\left\{1-\frac{(s^2-k^2-q^2)^2}{4k^2q^2}\right\}^{-1/2}\left(1-\frac{s^3}{(s^2+1)^{3/2}}\right)\nonumber\\
+\frac{32}{\pi}\int^{\infty}_{0}\frac{dkk}{(k^2+1)^{3/2}}
\int^{\infty}_{0}dq\left(q^2-\frac{q^3}{\sqrt{q^2+1}}-\frac{1}{2}\right)\nonumber\\
\times\int^{k+q}_{|k-q|}dss\left\{1-\frac{(s^2-k^2-q^2)^2}{4k^2q^2}\right\}^{-1/2}\left(1-\frac{s}{\sqrt{s^2+1}}\right)\nonumber\\
-\frac{32}{\pi}\int^{\infty}_{0}\frac{dkk}{\sqrt{k^2+1}}
\int^{\infty}_{0}\frac{dqq}{\sqrt{q^2+1}}
\int^{k+q}_{|k-q|}dss\left\{1-\frac{(s^2-k^2-q^2)^2}{4k^2q^2}\right\}^{-1/2}\nonumber\\
\times\frac{s}{\sqrt{s^2+1}}\frac{{\tilde{f}}^2(k,q,s)}{k\sqrt{k^2+1}+q\sqrt{q^2+1}+s\sqrt{s^2+1}}.
\end{align}
The results of numerical calculations of these integrals are presented in main text.

\newpage

\end{document}